# Classification of Topological Phase in 2D Photonic Continuous Media Using Electromagnetic Parameters

Xin-Tao He, Shuo-Shi Zhang, Xiao-Dong Chen and Jian-Wen Dong[*]

School of Physics & State Key Laboratory of Optoelectronic Materials and Technologies,
Sun Yat-sen University, Guangzhou 510275, China
* Email: dongjwen@mail.sysu.edu.cn

**Refractive index is a fundamental electromagnetic (EM) parameter that can describe photonic continuous media (PCM) traditionally as either transparency or opacity. Recently, topological theory offers a new set of phases to characterize PCM as either trivial or nontrivial, by using topological invariant which are not direct to EM parameters. As all of optical properties in PCM should be related to EM parameters, we formulate a topological index based on EM parameters and establish its phase map in this work. The map can analytically describe the deterministic condition for a topologically nontrivial phase. Our findings indicate that the topology of 2D bi-anisotropic PCM is determined by the sign of the topological index. Another EM parameter of pseudo surface impedance is also introduced for the opaque regions of PCM, showing that the topological opacity has a full range of impedance values ranging from negative to positive, while the trivial case only has either negative or positive impedance. The simulation results show that an interface between two opacities with differing index signs can support robustly optical propagation of topological edge states. As the index only depends on EM parameters, it will pave an insightful way to further understand the intrinsic properties of photonic topology.**

Topology, a concept originating from mathematics, has attracted widespread attention in the physics world, from condensed matter to classical waves [1, 2]. Recently, rapid developments in topological photonics have created an advanced platform at both microwave and optical frequencies for flexibly controlling various types of topological phases, as well as provided a new paradigm for designing on-chip nanophotonic devices [3-23]. These intriguing features in principle are derived from the topological band theory, which can theoretically predict nontrivial evolution of band structures and protections of topological states in the $k$ space [24]. For photonic systems, a general method to mimic the effective model of topological band theory is the exploration of photonic periodic lattice (PPL). As one of the common cases, photonic crystals support Bloch waves in its building blocks [17, 25], and some of the eigenmodes around high-symmetry $k$ points can be approximately described by the effective Hamiltonian. Based on topological band theory, one can use topological invariant to characterize topological phases as either non-trivial or trivial.

Photonic continuous media (PCM) are a class of optical homogeneous materials, such as metamaterials which are homogenized as effective media. As well known, most of metamaterials or artificial PCM can be described by a set of electromagnetic (EM) parameters [26, 27], i.e. permittivity $\varepsilon$ and permeability $\mu$. Based on the sign of $\varepsilon$ and $\mu$, such PCM have achieved lots of exotic EM responses which are difficult to be realized by natural existence of optical materials, including ultrahigh-index permittivity ($\varepsilon \gg 1$, $\mu > 0$), negative index ($\varepsilon < 0$, $\mu < 0$), magnetic mirrors ($\varepsilon > 0$, $\mu < 0$) and zero refractive index ($n = \sqrt{\varepsilon\mu} \approx 0$). The exploration of topologically non-trivial phases also extends to PCM under the consideration of certain material dispersion, which is analogous to the topological bands of PPL. In this way, PCM could maintain variety of topological features, even though it is lack of Bloch-wave mechanism. For example, one-way waveguide were realized in an interface between a 2D magnetized plasma and an ordinary reflector [28, 29], which shows an evidence for the topological edge states in PCM.

Later, an efficient method was developed to calculate the topological invariant of dispersive magnetized plasma and give a direct characterization of topologically non-trivial phase [30]. It was also found that 3D PCM with chiral response could be associated with topological phase and Weyl point [31-33]. The concept of direction-dependent refractive index was developed to analyze the topological features of zero-refractive-index materials [34]. The retrieval of topologically non-trivial phases in these works is mostly derived from band theory. However, it is not straightforward for PCM, as all of optical responses should be associated with their EM parameters. This issue inspires us to seek a new approach to bridge the gap between topological invariant and EM parameters.

In this work, as the consequences that the nonzero topological invariant is derived from the singularity of Berry curvature, we employ EM parameters to define a topological index related to the condition of the singularity. Based on refractive index and topological index, a topological phase map (see Fig. 1) is constructed to demonstrate the topologically nontrivial phase in 2D PCM, which is determined by the sign of topological index. An example of PCM with Lorentz-like dispersion is given to show that the evolution of topological index is well consistent with band theory. An interface between two opacities of different index sign gives rise to gapless edge states against sharp bending. Such bulk-edge correspondence is also described by pseudo surface impedance of complete gap.

At first, we consider a bi-anisotropic PCM with the following constructive relations,

$$\vec{D} = \varepsilon_0 \ddot{\varepsilon}_r \vec{E} + (\ddot{\chi}/c)\vec{H}, \vec{B} = \mu_0 \ddot{\mu}_r \vec{H} + (\ddot{\chi}/c)\vec{E}, \tag{1}$$

where $\ddot{\varepsilon}_r = \begin{pmatrix} \varepsilon_p & 0 & 0 \\ 0 & \varepsilon_p & 0 \\ 0 & 0 & \varepsilon_z \end{pmatrix}$, $\ddot{\mu}_r = \begin{pmatrix} \mu_p & 0 & 0 \\ 0 & \mu_p & 0 \\ 0 & 0 & \mu_z \end{pmatrix}$, $\ddot{\chi} = \begin{pmatrix} 0 & -i\xi & \\ i\xi & 0 & \\ & & 0 \end{pmatrix}$, (2)

are the relative permittivity, relative permeability and bi-anisotropic tensors, respectively. In general, optical materials respond differently for electric field and magnetic field due to $\ddot{\varepsilon}_r \neq \ddot{\mu}_r$. In order to construct a pair of pseudo spin degenerating, one of the solutions is to retrieve electromagnetic-dual parameter $\ddot{\varepsilon}_r = \ddot{\mu}_r = diag(\varepsilon_p, \varepsilon_p, \varepsilon_z)$. Thus the Maxwell equations can be divided into two non-relativistic equations for two sets of decoupled vectors $\vec{\psi}^\pm = \left(\sqrt{\varepsilon_0}E_x \mp \sqrt{\mu_0}H_x \quad \sqrt{\varepsilon_0}E_y \mp \sqrt{\mu_0}H_y \quad \sqrt{\varepsilon_0}E_z \pm \sqrt{\mu_0}H_z\right)^T$, where the superscript + (−) represents the spin-up (spin-down) eigenstate for in-phase (anti-phase) field pattern between $E_z$ and $H_z$. The detailed derivation can be seen in Appendix A of Supplemental Material. Note that all of 'spin' in this work represent pseudo spin.

With a simple deduction, the Maxwell equations can be generalized to be Schrödinger-like formation $\ddot{H}^\pm \vec{\psi}^\pm = \frac{\omega}{c} \vec{\psi}^\pm$ [see Appendix B of Supplemental Material]. The characteristic matrix $\ddot{H}^\pm$ is the effective Hamiltonian of PCM, which is similar to the role of Hamiltonian in quantum system. The eigenfields in PCM can be simplified as plane-wave form that $\vec{\psi}^\pm = \vec{\varphi}^\pm \exp(i\vec{k}\cdot\vec{r})$, where $\vec{\varphi}^\pm$ is the complex amplitude of the spin-polarized states. After solving the eigenvalues of effective Hamiltonian, the bulk dispersion is given as,

$$k_x^2 + k_y^2 = k^2 = \left(\frac{\omega}{c} n_{eff}\right)^2, \tag{3}$$

where $n_{eff} = \sqrt{\varepsilon_z(\varepsilon_p^2 - \xi^2)/\varepsilon_p}$ is the effective refractive index. Eq. (3) is the characteristic equation for both spin–up and –down modes, since they degenerate simultaneously in the bulk. Consider the BI PCM with Lorentz-like dispersion as the following form,

$$\varepsilon_p = \mu_p = 1 + \frac{\omega_A^2}{\omega_{op}^2 - \omega^2}, \quad \xi = \frac{\omega_B \omega}{\omega_{ok}^2 - \omega^2}, \quad \varepsilon_z = \mu_z = 1 + \frac{\omega_C^2}{\omega_{oz}^2 - \omega^2}, \tag{4}$$

where $\omega_{op}$, $\omega_{ok} = \omega_{op}$, $\omega_{oz} = 0.8\omega_{op}$, are the oscillation frequencies for different components, while $\omega_A = 1.5\omega_{op}$, $\omega_B = 0.75\omega_{op}$, $\omega_C = 0.9\omega_{op}$ represent the resonance strength, respectively. Here all parameters are normalized by the in-plane oscillation frequency $\omega_{op}$. Applying Eq. (4) to Eq. (3), we have analytic solution of the dispersion relation as shown in Fig. 2(a). The pink lines represent the band dispersions of two spin-degenerate modes. To subsequent discussion we refer to these four bands as bands 1-4, respectively. Separated by these bulk bands, there are three cyan gaps as the following frequency intervals: $\omega_{oz} < \omega < \omega_{op}$ for gap I, $\sqrt{\omega_{oz}^2 + \omega_C^2} < \omega < \left[ -\omega_B + \sqrt{\omega_B^2 + 4(\omega_{op}^2 + \omega_A^2)} \right]/2$ for gap II and $\sqrt{\omega_{op}^2 + \omega_A^2} < \omega < \left[ \omega_B + \sqrt{\omega_B^2 + 4(\omega_{op}^2 + \omega_A^2)} \right]/2$ for gap III.

To study the topological properties of those bulk bands and gaps, we should attain the Berry information, including the Berry connection $\vec{A}^\pm$, Berry curvature $\Omega^\pm$ and the corresponding spin Chern number $C_s$. By using the generalized method (Appendix C of Supplemental Material) and nonlocal approximation (Appendix D of Supplemental Material), the Berry connection $\vec{A}^\pm$ and the Berry curvature $\Omega^\pm = \nabla_k \times \vec{A}^\pm$ in PCM can be calculated analytically. Based on the integration of Berry curvatures over entire $k$ space, we have a quantized invariant to define the topology of spin–up and –down channels,

$$C^\pm = \frac{1}{2\pi} \iint \Omega^\pm dk_x dk_y . \qquad (5)$$

The net Chern number always vanishes (i.e. $C^+ + C^- = 0$) due to time-reversal invariance. The topological phase of the overall system can be characterized by another invariant, i.e. spin Chern number $C_s = (C^+ - C^-)/2 = C^+$. Note that the smooth distribution of Berry curvature satisfies Stokes' theorem: $C_s = \frac{1}{2\pi} \iint \Omega^+ dk_x dk_y = \frac{1}{2\pi} \oint_{k=\infty} \vec{A}^+ \cdot d\vec{l}$ [bands 1 and 2 in Fig. 2(b)] and thus the surface integral of Berry flux can be replaced by the line integral of Berry gauge field over the $k$-space boundary ($k = \infty$). The well-defined effective Hamiltonian ensures the Berry connection to be vanish at $k = \infty$, so that the line integral of equifrequency contour of infinite wavevector attains $C_s = \frac{1}{2\pi} \oint_{k=\infty} \vec{A}^+ \cdot d\vec{l} \equiv 0$. To obtain nonzero $C_s$, the singularity of Berry curvature should be found. In band 3 (4) of Fig. 2(b), we can clearly observe singular peak of spin-up Berry curvature with anticlockwise (clockwise) vortex of Berry connection. The singular vortex indicates a source of Berry gauge field in accordance with the fact that $C_s$ is -1 (+1). In these two cases, the Strokes' theorem can be recovered by adding a summation of line integrals in the vicinity of singular *k* points, i.e. $C_s = \frac{1}{2\pi}\oint_{k=\infty}\vec{A}^+ \cdot d\vec{l} - \frac{1}{2\pi}\sum_{k_i}\oint_{R^+}\vec{A}^+ \cdot d\vec{l}$, where $R^+$ is a circle in the vicinity of singular $\vec{k}_i$ point. The summation term provides nonzero value of spin Chern number and thus leads to a topological phase transition. In other words, to guarantee the existence of topological nontrivial phase, it is more significant to explore the singularities of Berry curvature.

In fact, regardless of band theory, the optical properties of PCM should be related to its EM parameters. This issue inspires us to bridge the gap between topological phase and EM parameters. We start with a fundamental parameter, i.e. the square of effective refractive index (SERI) $n_{eff}^2 = \varepsilon_z\left(\varepsilon_p^2 - \xi^2\right)/\varepsilon_p$. Figure 3(a) show the SERI of bi-anisotropic PCM as a function of frequency. The transport properties are traditionally determined by the sign of SERI. For $n_{eff}^2 > 0$ (pink bands), the transparent waves propagate in the bulk media freely, while the opaque modes localize in the surface and exponentially decay into the bulk when $n_{eff}^2 < 0$ (cyan gaps). As SERI can't characterize the topology of PCM, we should seek a new index that can describe the relationship between topological phase and EM parameters.

As stated above, the transition of topological phase is related to the singularity of Berry curvature. The expression of Berry curvature can be analytically written as

$$\Omega^\pm = \mp 2\left[\left(\xi + \omega\partial_\omega\xi\right)\left(\varepsilon_p^2 + \xi^2\right) - 2\xi\varepsilon_p\left(\varepsilon_p + \omega\partial_\omega\varepsilon_p\right)\right]/\left[W_0 k_0^2\left(\varepsilon_p^2 - \xi^2\right)^2\right], \tag{6}$$

where $W_0$ is proportional to the time-averaged energy density that $W_0 > 0$. To observe the distribution of Berry curvature and retrieve topologically nontrivial phase, all of the following discussions focus on nonzero BI coefficient. According to Eq. (6), the Berry curvatures go to infinity when the term $\varepsilon_p^2 - \xi^2$ approaches to zero and thus leads to a transition of topological phase. In other words, we can use such index to characterize the topology of PCM. Figure 3(b) plots the topological index $\gamma = \varepsilon_p^2 - \xi^2$ as a function of operation frequency. Two special points ($\varepsilon_p^2 - \xi^2 = 0$) are highlighted as blue dots. They guarantee the existence of the singularity of Berry curvature which leads to topological transition. Below point A, including bands 1-2 and gaps I-II, the material attributes to trivial topology with topological index $\varepsilon_p^2 - \xi^2 > 0$. At the intersection between gap II and band 3, the topological index experiences from positive to negative value. The change of sign of topological index leads to the singular structure of Berry curvature and provides a topological transition from trivial to nontrivial phase. Therefore, the region with negative topological index carry topologically nontrivial properties. Similarly, another topological transition can be also observed around point B.

A topological phase map is defined to combine the information of SERI and topological index, as shown in Fig. 1. The phase map can be divided into four quadrants, i.e. ordinary transparency, topological transparency, topological opacity and ordinary opacity. There are two types of topological transition highlighting by red arrows. As SERI $n_{eff}^2 = \varepsilon_z \left( \varepsilon_p^2 - \xi^2 \right) / \varepsilon_p$ include the term of topological index $\varepsilon_p^2 - \xi^2$, the topological transition points ($\varepsilon_p^2 - \xi^2 = 0$) will inevitably appear at the origin of phase map. In other words, the transition of topology will always be accompanied by transparent-opaque transition. Figure 3(c) gives the evolution of phase map in bi-anisotropic PCM, as the operation frequency increases continuously. The material response and numerals labels are identical to Fig. 2(a). The medium first behaves as ordinary transparency (band 1) and then turns to be ordinary opacity (gap I) with a resonance of $\varepsilon_z$. As the increasing of operation frequency, the optical response jump back to the first quadrant (band 2) and passes through the horizontal axis without resonance. We should note that all of the above trajectories locate at the topologically trivial semi-space. And then, two types of topological transition are found near the origin of coordinate: the first one is from ordinary opacity to topological transparency [i.e. point A in Fig. 3(b)], while the second one is between topological opacity and ordinary transparency [i.e. point B in Fig. 3(b)].

On the other hand, the surface impedance is related to the existence of interface states. Inspired by this method, we introduce the concept of surface impedance into the bi-anisotropic PCM. Figure 4(a) shows the definition of pseudo surface impedance on the downward side of the boundary as the ratio of the *x*-direction spin wave to the *z*-direction spin wave, yielding $Z_s^{\pm} = \psi_x^{\pm}(y=0^-)/\psi_z^{\pm}(y=0^-)$, where $y = 0$ defines the boundary. Based on Eq. (12) of Supplemental Material, we can analytically obtain the pseudo surface impedance in theory. Here, we only focus on the pseudo surface impedances in the opaque regions, whose real part vanish in the complete gap. The imaginary part of pseudo surface impedances for spin-up states are presented in Fig. 4(b). These three different PCMs have the same EM parameters to Fig. 2 except for normalized in-plane oscillation frequency. Medium 1 preform as a topologically trivial opacity in the gap region with negative impedance, while medium 2 possesses positive impedance in the topologically trivial opaque region. The topologically non-trivial opacity (medium 3) experiences full-range impedance from positive infinity to negative infinity. To guarantee the existence of an interface state, one should retrieve matched impedances on each side of interface, i.e. $Z_{s1} + Z_{s2} = 0$. For example, a traditional method is to form an interface between positive-impedance opacity and negative-impedance opacity. However, some certain frequencies cannot verify the impedance-matched condition, such that there is lack of surface waves. The topologically nontrivial gap, with full-range impedance from positive infinity to negative infinity, gives a robust mechanism to match another impedance of trivial gap. This is another demonstration why the gapless dispersion of surface waves can be formed at the interface between nontrivial gap (opacity) and trivial gap (opacity).

Next, we will focus on the propagation of edge states forming by two diverse opacities. To get a better understanding of spinful edge states, we construct a one-dimensional interface along $x$ axis formed between two semi-infinite materials. The edge dispersion for spin-up and -down can be presented as follows,

$$\pm k_x \frac{\xi_1}{\varepsilon_{p1}^2 - \xi_1^2} + \frac{\varepsilon_{p1}\sqrt{k_x^2 - k_0^2\left(\varepsilon_{p1}^2 - \xi_1^2\right)\varepsilon_{z1}/\varepsilon_{p1}}}{\varepsilon_{p1}^2 - \xi_1^2} = \pm k_x \frac{\xi_2}{\varepsilon_{p2}^2 - \xi_2^2} - \frac{\varepsilon_{p2}\sqrt{k_x^2 - k_0^2\left(\varepsilon_{p2}^2 - \xi_2^2\right)\varepsilon_{z2}/\varepsilon_{p2}}}{\varepsilon_{p2}^2 - \xi_2^2} \quad (7)$$

where the subscripts 1 and 2 represent the materials on each side of interface. After solving Eq. (7), we can obtain the edge dispersion at an interface between an $C_s = 0$ ordinary opacity (medium 1) and a $C_s = -1$ topological opacity (medium 3) in Fig. 5(a). As the time-reversal partners, the dispersions for spin-up (blue) and spin-down (red) edge states are symmetrical with respect to the plane of $k_x = 0$, and gaplessly cross over the complete gap. As a consequence, the rightward unidirectional propagation can be excited by a spin-up point source [blue vortex in Fig. 5(b)], as expected from the bulk topology. Such spin-momentum locking property is the photonic analog of quantum spin Hall effect. Figures. 5(c-f) show the simulation results (calculated by COMSOL Multiphysics) of optical propagation along the two interfaces, when the operation frequency of all excited source is $\omega = 0.9\omega_{op}$. The topological edge state can be robust against the 'SYSU' shaped bending, as depicted in Fig. 5(c). The incident light couple to rightward wave and pass through the bending interface without backscattering, even though using a spinless $E_z$-polarization source (black dot). For comparison, we also give a control case of gapped edge dispersion supported by an interface between two ordinary opacities (medium 1 and medium 2). Figs. 5(e) and 5(f) have the same numerical setup to the left panel, except for topologically trivial interface. Due to lack of topological protection, the interface wave is failure against both unidirectional excitation and robust propagation.

In summary, we have successfully applied a topological index based on EM parameters to demonstrate the topologically nontrivial properties in a class of 2D PCM with bi-anisotropic response.

With an analytic deduction, we have concluded that the topology is determined by the sign of topological index. An example with Lorentz-like dispersion is given to verify this key point, including the evolution of topological index, pseudo surface impedance of bulk-edge correspondence, and topologically edge states. Since such index only depends on EM parameters, it will benefit to simplify the exploration of topological metamaterial, and will lead to novel fundamental physics and device applications in the field of metamaterial. Moreover, our method is compatible to analyze the topology of magnetized plasma [29, 30], and the bi-anisotropic metamaterials have been constructed by EM-dual 'meta-atom' between two metal plates. [35, 36].

**Figures and Figure Captions**

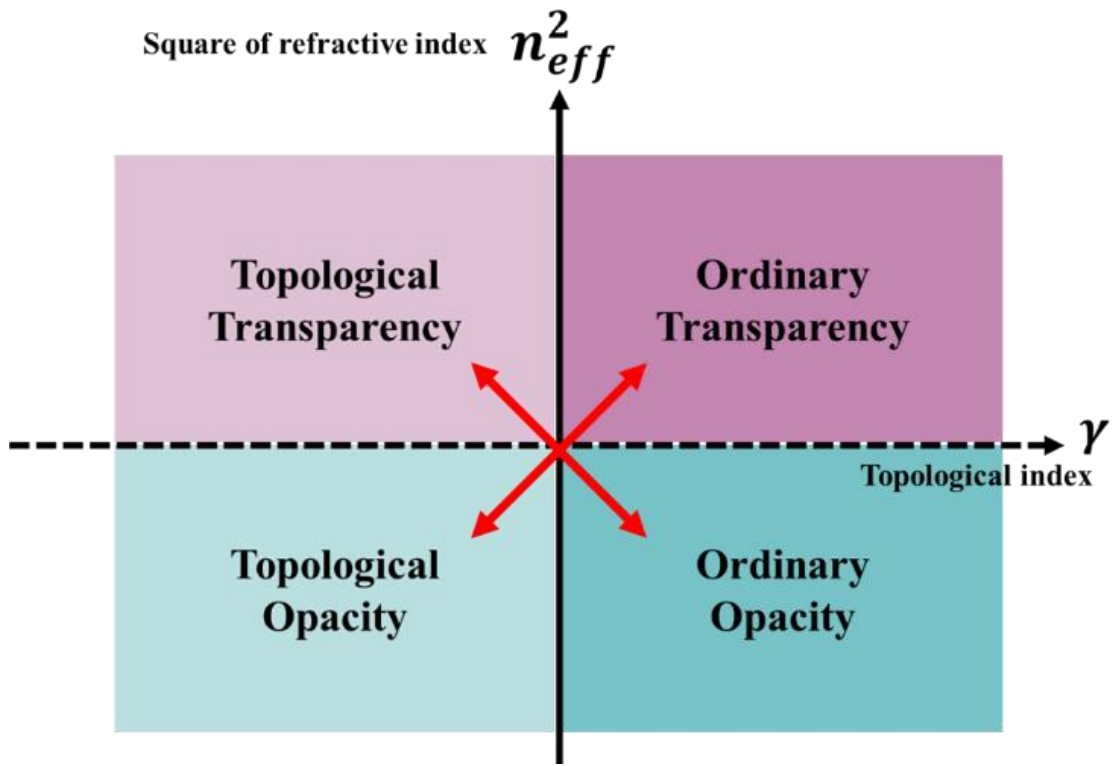

FIG. 1. Topological phase map of photonic continuous media (PCM), which is divided into four quadrants, i.e. ordinary transparency, topological transparency, topological opacity and ordinary opacity. Red arrows indicate two types of topological transition.

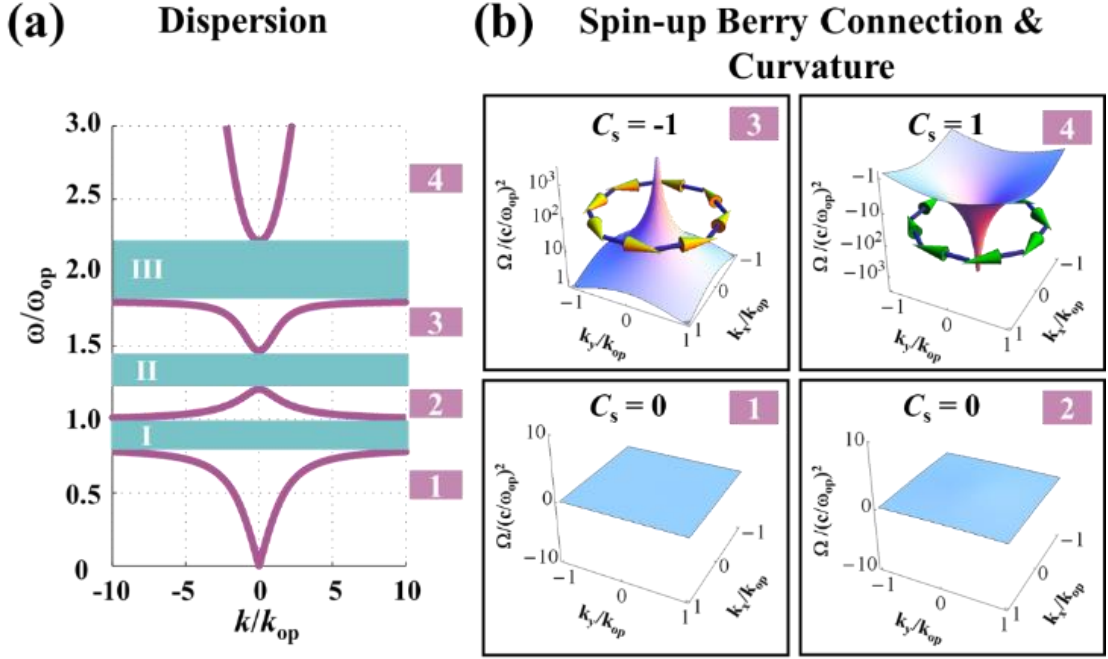

FIG. 2. (a) Dispersion relation of electromagnetic-dual bi-anisotropic PCM. $\omega_{op}$ is the in-plane oscillation frequency. The pink lines represent the band dispersions of two spin-degenerate bulk modes (referred as bands 1-4), while there are three cyan gaps I-III separated by these bands. (b) Spin-up Berry curvature $\Omega^+$ and schematic view of spin-up Berry connection $\vec{A}^+$ for bands 1-4. For band 3 (4), we can clearly observe the singular peak (dip) of Berry curvature at the center of $k$ space, which is accompanied with anticlockwise (clockwise) vortex of Berry connection.

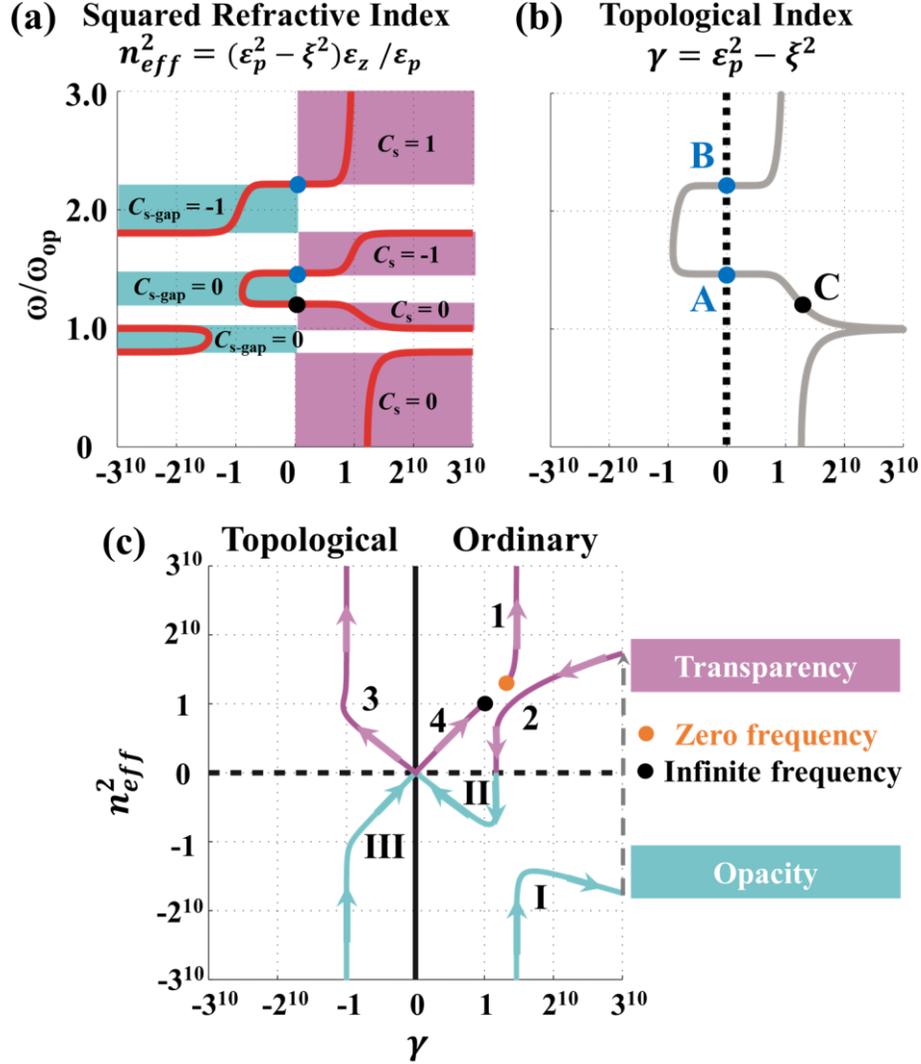

FIG. 3. (a) The square of effective refractive index (SERI) $n_{eff}^2 = \varepsilon_z\left(\varepsilon_p^2 - \xi^2\right)/\varepsilon_p$ as a function of frequency. (b) Topological index $\gamma = \varepsilon_p^2 - \xi^2$ as a function of frequency. Two zero-topological-index points ($\varepsilon_p^2 - \xi^2 = 0$) are highlighted as blue dots. (c) Evolution of topological phase map as the operation frequency increases continuously. The material response and numerals labels are identical to Fig. 2(a).

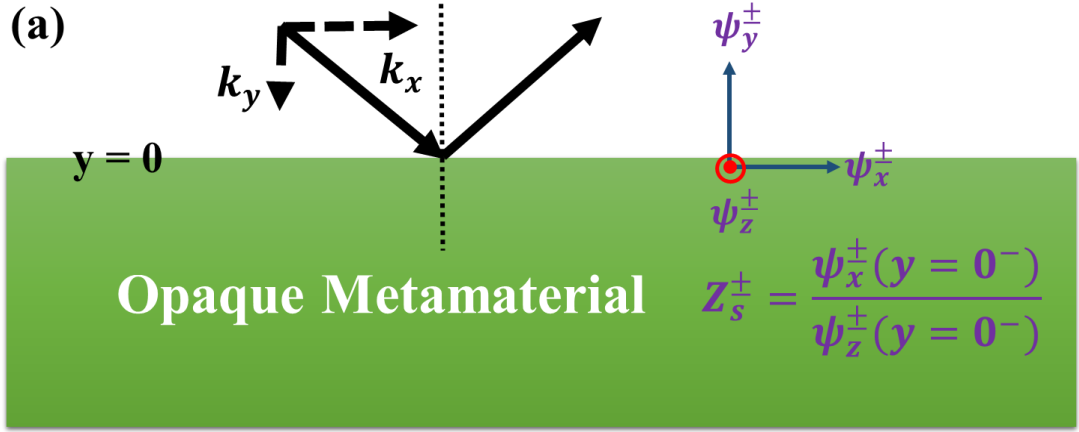

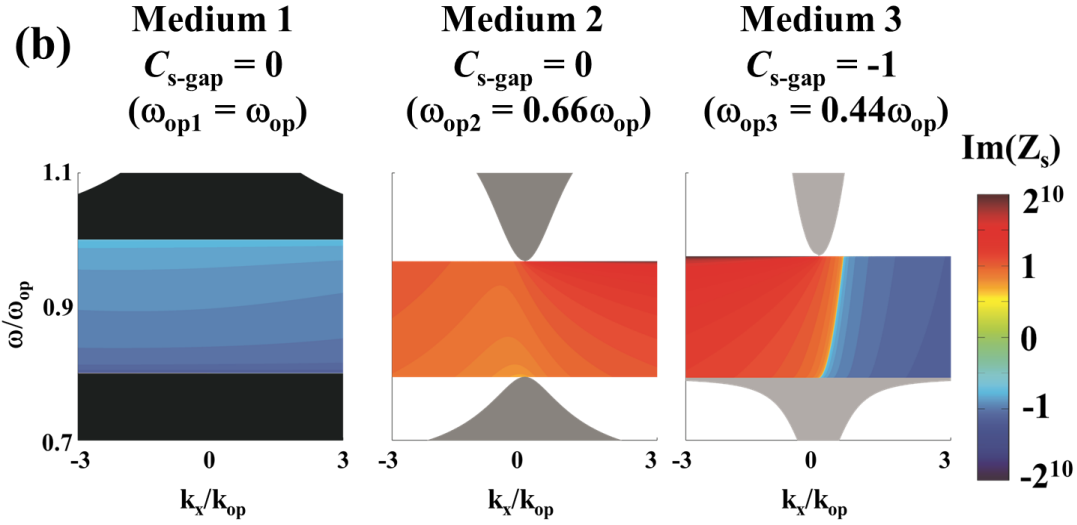

FIG. 4. (a) Definition of pseudo surface impedance as the ratio of the *x*-direction spin wave to the *z*-direction spin wave. (b) Imaginary part of surface impedance for spin-up states in three different PCM, as the same EM parameters to Figure 2 except for normalized in-plane oscillation frequency. For simplicity, the pseudo surface impedances are only given in the opaque regions of each medium.

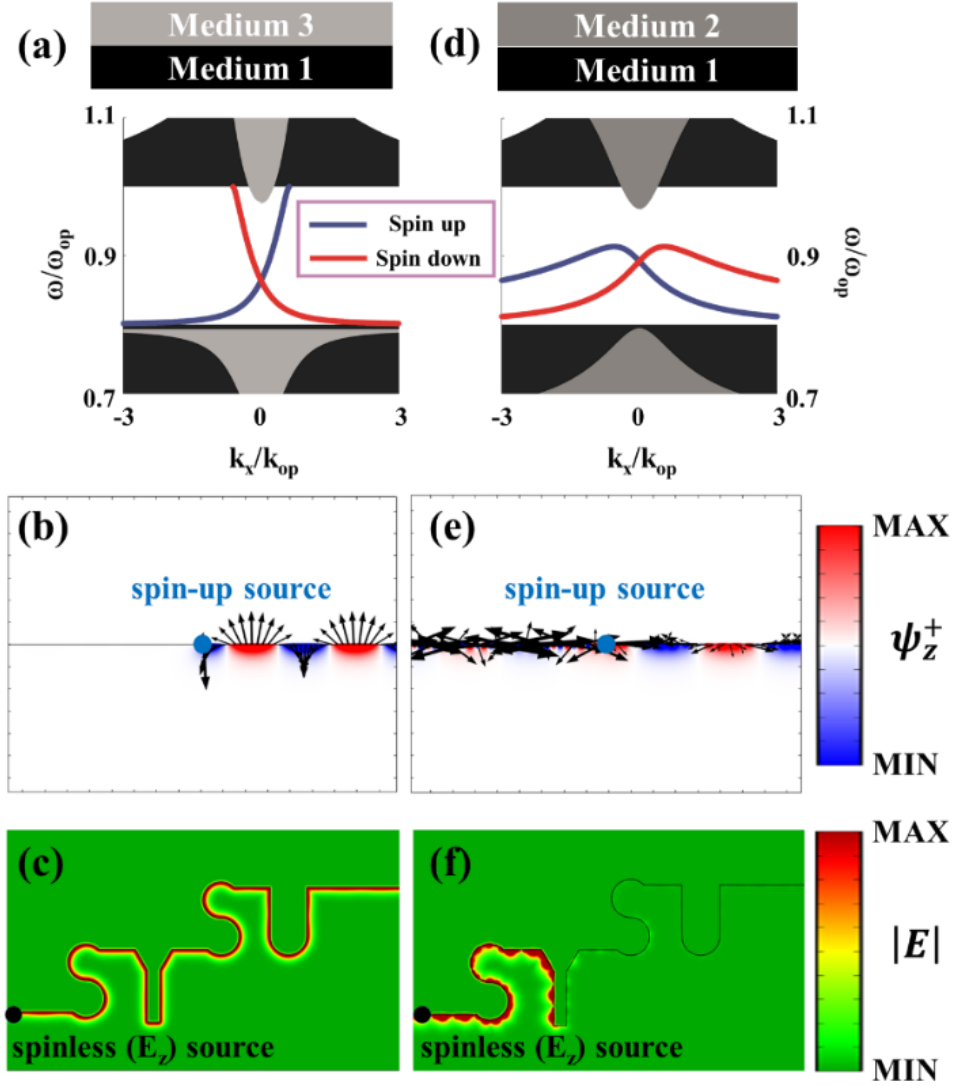

FIG. 5. (a) Edge dispersion at an interface between an $C_s = 0$ ordinary opacity (medium 1) and a $C_s = -1$ topological opacity (medium 3). (b) Rightward unidirectional couple excited by a spin-up point source (blue dot). The black arrows represent the in-plane components ($\psi_x$, $\psi_y$) of edge states. (c) Robust propagation of topological edge state against the 'SYSU' shaped bending, by using a spinless $E_z$-polarization source (black dot). (d) Edge dispersion supported by an interface between two ordinary opacities (medium 1 and medium 2). (e, f) Same numerical setup to the left panel, except for topologically trivial interface.